\documentclass[twocolumn,
superscriptaddress,
showpacs,%preprintnumbers,
 amsmath,amssymb,
 aps,
 pra,
floatfix,
]{revtex4}

\usepackage{graphicx}% Include figure files
\usepackage{dcolumn}% Align table columns on decimal point
\usepackage{bm}% bold math
\usepackage[T1]{fontenc}
\usepackage{lmodern}

\newcommand{\mum}{\mu {\rm m}}
\newcommand{\xx}{\bf x}
\newcommand{\ba}{\bf a}
\newcommand{\kk}{\bf k}
\newcommand{\eq}[1]{(\ref{eq:#1})}

\begin{document}

\preprint{APS/123-QED}

\title{The study of random vorticity in quantum fluids through interference fluctuations}% Force line breaks with \\
%\thanks{A footnote to the article title}%

\author{M.~Wouters}%
\affiliation{TQC, Departement Fysica, Universiteit Antwerpen, Belgium}

\date{\today}% It is always \today, today,
             %  but any date may be explicitly specified

\begin{abstract}
We study the vortex dynamics of a quantum degenerate Bose gas through the intensity fluctuations of the interference from particles extracted at two different positions. It is shown numerically with classical field simulations that an interacting Bose gas with proliferating vortices exhibits long correlation times for these intensity fluctuations. This behavior is contrasted with the case of a noninteracting gas, that we describe analytically, and with the case of a well condensed Bose gas without vortices. We discuss the observability of our predictions in quantum fluids of exciton-polaritons.
\end{abstract}

\pacs{67.10.Ba, % boson degeneracy in quantum fluids
03.75.Lm,				% vortex dynamics in Bose-Einstein condensates
71.36.+c  %polaritons
}% PACS, the Physics and Astronomy
                             % Classification Scheme.

%\keywords{Suggested keywords}%Use showkeys class option if keyword
                              %display desired
\maketitle

\section{Introduction}
Quantized vortices play a central role in the thermal and dynamical properties of quantum degenerate Bose gases. Vortex lattices in rotating Bose-Einstein condensates \cite{vortlatt},  spontaneous vortices above the Berezinskii-Kosterlitz-Thouless (BKT) transition \cite{bkt_dalibard}, discrete vorticity in turbulent superfluids~\cite{turb_sup} and Kibble-Zurek (KZ) vortices \cite{anderson} are prominent examples of the importance of quantized vortices.
 
Their experimental visualization relies either on the associated density minimum or the singularity in the phase profile. Where the density is usually the more easily accessible observable, only a measurement of the phase gives unambiguous proof of the quantized circulation around the vortex core. Vortices in both Bose-Einstein condensates of ultracold atomic gases and of exciton polaritons in semiconductor microcavities have been detected in this way. To this end, the particles coming from different regions in the same condensate (of exciton-polaritons \cite{konst_vort}) or from independent condensates (of atomic gases \cite{ketterle_int,bkt_dalibard}) were interfered. A dislocation in the interference pattern proves the presence of a quantized vortex. With this technique, vortices in rotating atomic condensates have been evidenced \cite{ketterle_int}, as well as spontaneous vortices above the transition temperature of a 2D atomic Bose gas \cite{bkt_dalibard}. In exciton-polariton condensates, quantized vortices induced by the interplay of pumping and decay with a disordered potential landscape were discovered \cite{konst_vort}.

Regarding their observation, the distinction between deterministic and spontaneous vortices is essential. The former occur on positions that are determined by the system geometry -- examples are the vortex lattices of rotating condensates or the vortices of exciton-polaritons pinned to disorder. Their predictable positions makes that there are no restrictions on the time it takes to record the interferrogram. For the observation of the spontaneous quantized vortices on the other hand, it is essential that the experiment can be performed on a time scale that is fast with respect to the vortex motion. For experiments with ultracold atoms, where the intrinsic time scales are of the order of ms, this poses no problem. The spontaneous BKT \cite{bkt_dalibard} and KZ \cite{anderson} vortices have indeed been successfully detected.
For exciton-polariton condensates \cite{iac_rev}, that are a billion times faster (typical time scale is 1 ps), the required detection speed poses a technological challenge that is yet to be overcome. The time it takes to record an interferrogram is much longer than the time scale on which the vortices move. As a consequence, after the temporal averaging, no phase defects can be observed, but only a decreased spatial coherence. This difficulty has made spontaneous hydrodynamic vortices under cw excitation \cite{amo_soliton} and possible KZ vortices \cite{nardin} elusive and forms a limitation for the experimental study of quantum turbulence, suggested in Ref. \cite{berloff_turb}.

While no image of the polariton interferrograms can be recorded on a ps time scale, it is possible to study its intensity fluctuations with a standard Hanburry Brown and Twiss setup \cite{kasprzak_g2}. We will propose in this paper to study these intensity fluctuations in order to characterize the vorticity in the fluid. The idea is presented in Sec. \ref{sec:int}. Where the intensity fluctuations can be computed analytically in the noninteracting case, numerical classical field simulations are presented in Sec. \ref{sec:class} to investigate the interacting case. The effect of a finite particle life time, relevant for experiments with exciton-polaritons is addressed as well. We draw conclusions in Sec. \ref{sec:concl}.

\section{Interference intensity fluctuations \label{sec:int}}

\begin{figure}[tb]
\includegraphics[width=0.45\textwidth]{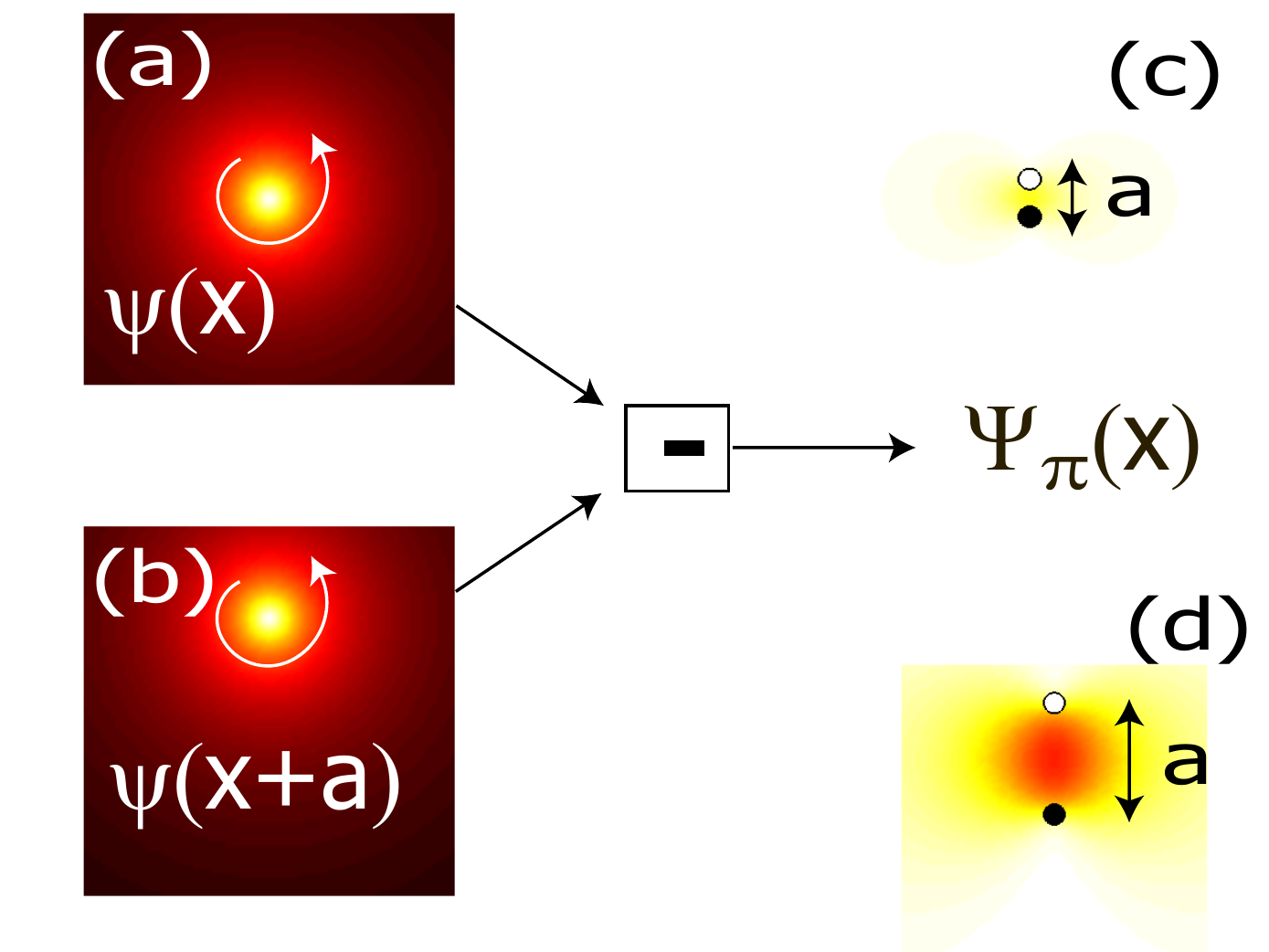}% Here is how to import EPS art
\caption{Sketch of the density that results from interfering a condensate containing a vortex (a) with a displaced copy (b), with a phase shift of $\pi$ between the two. The interferrogram shows a density maximum between the vortex cores, whose size is determined by the displacement $a$. Panels (c) and (d) show two distances $a$, that are smaller and larger than the vortex core size respectively. }
\label{fig:g2I_sketch}
\end{figure}

For a schematic understanding of our proposal, we have sketched in Fig.~\ref{fig:g2I_sketch} an interferrogram of a condensate containing a vortex with a displaced copy of itself. Due to a $\pi$ phase shift between the two interfering beams, a non vanishing intensity is visible only between the positions of the vortices, that are marked with a full circle (vortex position on the condensate image) and an empty circle (vortex position on the displaced image). Panel (c) shows a separation of the interfered images $a$ that is smaller than the vortex core size $\xi$. The maximal density in the interferrogram is much smaller in this case than in panel (d), where the separation $a$ is larger than the vortex size. The reason is simply that for small separation, constructive interference occurs only in the vicinity of the vortex where the density is suppressed.

It is expected that vortices that move ballistically should give rise to correlations in the density fluctuations over a time that scales as $\tau_B \sim a$. In the case of diffusive vortex motion on the other hand, one rather expects $\tau_D \sim a^2$. Below, we will show analytically that this scaling constitutes a qualitative difference from the case of the noninteracting gas where the correlation time depends much weaker on $a$. The scaling $\tau_c \sim a, a^2$ in the presence of vortices will be addressed with numerical simulations in a classical field model.

%{\em Density fluctuations of the interferrogram --} 
When the light extracted at the positions $\xx$ and $\xx+\ba$ is interfered, the intensity incident on the detector is
$\hat \Psi_\theta=\hat \psi(\xx)+e^{i \theta} \hat \psi(\xx+\ba)$. When $\theta$ is chosen so to minimize the amplitude of $\Psi$, the intensity on the detector is the smallest and the relative fluctuation should be the largest. We restrict to this case in the following of the paper. For ordinary condensation in a $\kk=0$ state, this corresponds to $\theta=\pi$, but in the general case, a phase that suits the condensate wave function should be chosen. In typical experiments with polariton condensates \cite{kasprzak}, this means that the density fluctuations are measured at a minimum of the interference fringes.
The second order correlation function
\begin{equation}
g^{(2)}_I({\bf a} ,t,\tau) \equiv \langle \Psi^\dag_\pi(t) \Psi^\dag_\pi(t+\tau) \Psi_\pi(t+\tau) \Psi_\pi(t)  \rangle 
\label{eq:g2I}
\end{equation}
quantifies the correlations between a density fluctuation at time $t$ and one at time $t+\tau$.

When the fluctuations on top of the condensate are Gaussian, the density fluctuations of the interferrogram are Gaussian as well. We then find in the steady state a bunching of the fluctuations: $g^{(2)}_I({\bf a},t=0)=2g^{(2)}_I({\bf a},t=\infty) $.
When the different momentum components are assumed to have no phase relation, the decay time of the interference intensity fluctuation can be computed analytically by using Wick's theorem (see appendix)
\begin{equation}
g^{(2)}_I(a,\tau)=g^{(2)}_I(a,0) \frac{1+|f(\tau)|^2}{2},
\label{eq:g2I_noint}
\end{equation}
where $f(\tau)$ is 
\begin{equation}
f(\tau)={\mathcal N} \int d^D {\bf k} \; e^{-i \epsilon(k) \tau} n({\bf k}) \sin^2\left(\frac{{\bf k} \cdot {\bf a}}{2}\right),
\label{eq:f}
\end{equation}
where $\epsilon(k)$ is the single particle dispersion.
The normalization constant ${\mathcal N}$ is determined such that $|f(\tau=0)|=1$. Eq. \eqref{eq:g2I} expresses a simple relation between the density fluctuations in the interferrogram and the momentum distribution of the fluid.

\begin{figure*}[tb]
\includegraphics[width=\textwidth]{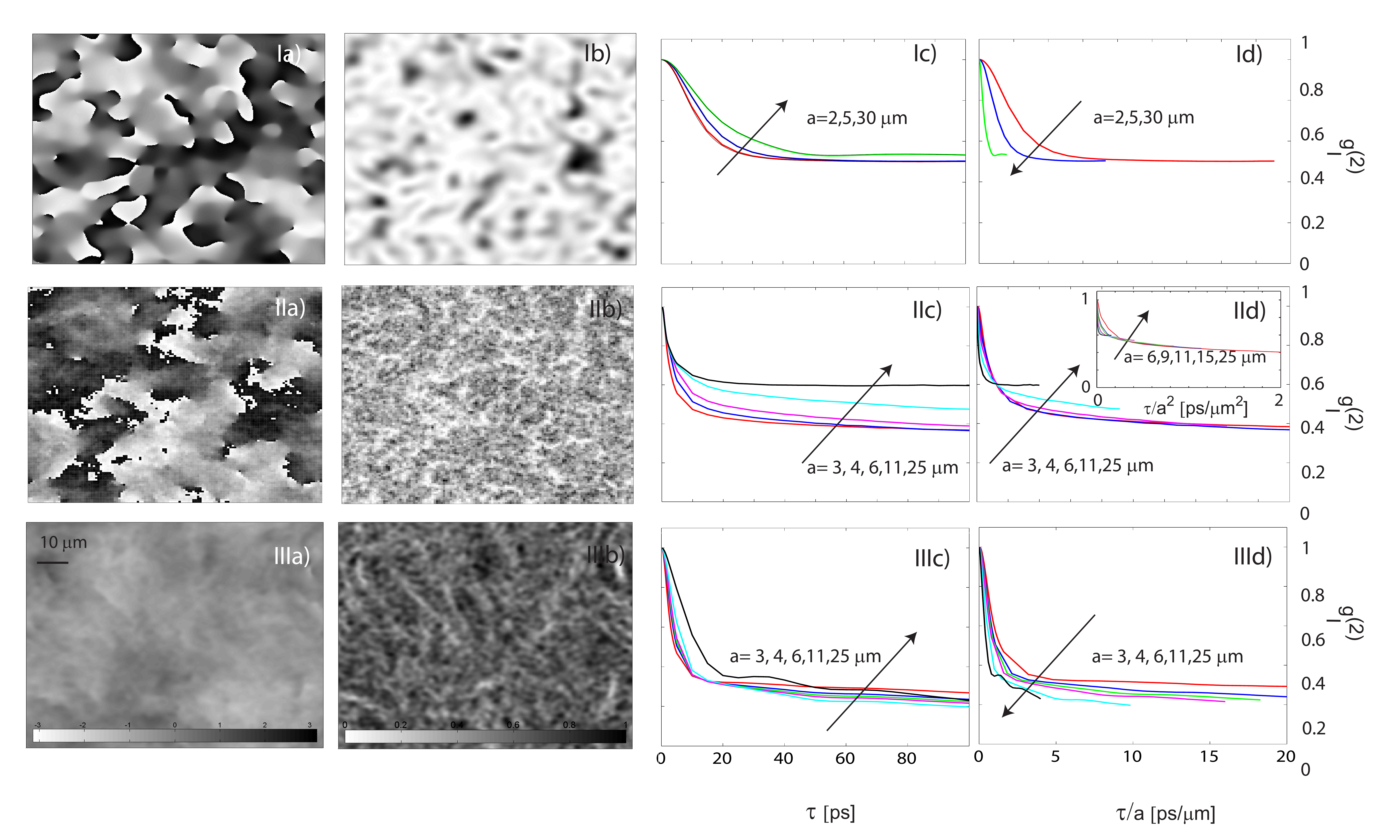}% Here is how to import EPS art
\caption{ Phase (a) and density (a) profile of a snap shot of the classical field that is evolved with the GPE for $t=50 {\rm ps}$ from a Gaussian momentum distribution (width $0.5 \mum^{-1}$) without interactions (I), with interaction strength $g n=0.5 {\rm meV}$ (II) and from a condensed state (96 percent condensate) with Gaussian occupation of higher momentum states (III). Panels (c) show the corresponding $g^{(2)}_I$ (normalized to its value at zero delay) as a function of the delay $\tau$ for several values of the separation $a$. The panels (d) show the same, but for a rescaled time variable $\tau/a$. The inset in panel IId) shows $g^{(2)}_I$ as a function of $\tau/a^2$.
Other parameters: effective mass $\hbar/m = 1 {\rm \mum^2 meV}$, interaction energy $g n=0.5 meV$.  Simulations were performed on a system of size 100x100 $\mu$m and 128x128 grid points.  }
\label{fig:results}
\end{figure*}
The validity of Eq. \eqref{eq:g2I_noint} depends on the absence of correlations between the different momentum components. This condition is satisfied for the non-interacting Bose gas.  When interactions are turned on, the momentum components of the gas become correlated, affecting the fluctuations of the interferrogram. We will study this case numerically in the next Section.

\section{Classical field simulations \label{sec:class}}
 We will investigate the weakly interacting Bose gas numerically based on a classical field model\cite{classfield}.
In this method \cite{goral}, the quantum field is replaced by a classical field whose dynamics is described by the Gross-Pitaevskii equation (GPE)
\begin{equation}
i \hbar \frac{d}{dt}\psi=\left(-\frac{\hbar^2}{2m}\nabla^2 + g|\psi|^2 \right) \psi.
\label{eq:GPE}
\end{equation}
Here, we have used a quadratic disperion with mass $m$ and $g$ is the strength of the contact interaction.
The evolution of the classical field with the GPE gives a reasonably accurate description of the thermal bose gas, including the BKT transition \cite{bisset}. In this work, we do not to focus on the spontaneous vortices of the BKT type. Actually in the the weakly interacting Bose gas, no well defined spontaneous vortices exist \cite{giorgetti}, due to the large magnitude of density fluctuations close to the transition temperature.

It is still possible to construct a weakly interacting quasi-condensate with well defined vortices by a suitable initial condition. Experimentally such a state can be the result of a sudden temperature quench of a gas of ultracold atoms \cite{berloff}, that produces vortices of the Kibble-Zurek type \cite{anderson}. In the polariton case, the initial state can be chosen at will, because under resonant excitation, it is directly imprinted by the excitation laser. 
A different approach is a quantum turbulent state where vorticity forms when the bose gas is driven externally \cite{bagnato,berloff_turb}.
In order to study a quasi-condensate with vortices, we choose a simple Gaussian initial condition in momentum space with random phases for each state $\psi(k) = \exp(-k^2/k_0^2 + i \eta(k))$, where $k_0$ determines the width of the momentum distribution and $\eta(k)$ is an uncorrelated uniform random number in the interval $\eta \in [0,2\pi[$. In the simulations presented in Fig. \ref{fig:results}, the value $k_0=0.5 \mu {rm}^{-1}$ was chosen, but we have checked that the results are not sensitive to this width.  The phase of such a state in real space has a number of singularities that increases with the width of the state in momentum space. In the absence of interactions, we can compare the results of our numerical simulations directly with \eq{g2I_noint}. This is shown in Figure \ref{fig:results} (Ic). The numerical simulations are consistent with our analytic calculations. 

%We wish to stress again that the numerical data show a correlation time that does not show any ballistic or diffusive scaling, even though the phase shows many dislocations [see Fig. \ref{fig:results} (Ia)] . The physical reason is that the phase singularities arise rather from interfering waves than from well defined vortex excitations.

When interactions are turned on, we numerically observe the expected narrowing of the momentum distribution, corresponding to the buildup of longer range spatial correlations. From the real space image, it appears that vortices become well defined objects with a characteristic size of a few $\mum$. This stands in stark contrast to the noninteracting case [compare in Fig.\ref{fig:results} panels (Ia,b) with panels (IIa,b)]. As discussed above, we do not expect well developed vortices in the equilibrium state. We therefore study the interference fluctuations in the transient. In our simulations we have observed that the change of $g^{(2)}_I$ in time does not change the qualitative features that we discuss below. 

The decay of density fluctuations in the interferrogram [see Fig. \ref{fig:results} (IIc)] is dramatically changed with respect to the noninteracting case. It now varies  now much more with the separation $a$ between the points that are interfered. The further the points are apart, the longer the correlation time of the fluctuations. To check quantitatively for a ballistic scaling $\tau \sim a$, we have rescaled in Fig. \ref{fig:results} (IId) the $\tau$-axis by $a$. The different curves taken at different displacements $a$ indeed overlap for the smaller distances ($a \leq 6 \mum$). 

Strikingly, the curves for larger separation $a$ in Fig. \ref{fig:results} IId) are above the ones for lower $a$. This indicates possible diffusive behavior of vortices at large distances \cite{minnhagen}. The inset in Fig. \ref{fig:results} IId) shows that a diffusive scaling $\tau\sim a^2$ works indeed well for larger separation. For the largest distances, there is again a slight deviation toward larger correlation times.

%Note that at short times, all curves in Fig. \ref{fig:results} IIc) decay at the same rate, independent of $a$. This is explained by the fact that at short times, the intensity fluctuations are dominated by the highest frequency single particle excitations, for which interactions can be neglected. Consequently, on the figure with rescaled time axis [panel IId)], the curves for larger separation $a$ are the lowest at short times. 

Comparing in Fig. \ref{fig:results} the panels (I) with the panels (II), it is clear that the observable $g^{(2)}_I$ is not simply related to the presence of dislocations in the phase profile. Dislocations in the phase profile occur both in the interacting and noninteracting case, but the phase dynamics is very different. Our interpretation of the different phase dynamics of the following. In the noninteracting case, the bose gas can be described in terms plane wave excitations. This is precisely the assumption that leads to the prediction \eqref{eq:g2I_noint}. Vortices, i.e. zeros of the wave function, are incidental and bear not much physical significance. The situation changes when interactions are introduced. Plane waves are then no longer the good excitations to characterize the system. This is witnessed very clearly by the dynamics of the interference that changes dramatically between panels (I) and (II). The dynamics shown in panels (II) can clearly not be explained by the prediction \eqref{eq:g2I_noint} for uncorrelated plane waves. In a real space picture, the correlations between the plane waves gives rise to the appearance of vortices with a characteristic size (the healingh length) as elementary excitations. 

These intensity fluctuations of the interferrogram thus give information that is complementary to the direct measurement of the condensate phase that identifies the zeros of the wave function. This observable could therefore be of interest for ultracold atoms as well. Experimentally, it could be measured by extracting atoms from two spatially separated regions in the condensate and counting them with a single atom counter, as in Ref. \cite{esslinger_g2}.

It is also interesting to investigate the case of a well condensed quantum fluid with small fluctuations on top. For the numerical study, we have taken an initial condition with $96$ percent of the particles in the condensate and a Gaussian distribution in momentum space for the thermal component. In this state with good spatial coherence, the fluctuations are too weak to induce free vortices. The Bogoliubov excitations are weakly occupied and can be considered to be a noninteracting gas. We thus expect the decay of $g^{(2)}_I$ to depend less on $a$.
Panel (IIIc) in Fig. \ref{fig:results}  shows that the dependence on $a$ is indeed much weaker than in panel (IIc). Especially when the time $\tau$ is rescaled with the separation $a$, this becomes very clear. Again, as in the noninteracting case, the curves with a longer separation decay faster in this graph than the ones with a shorter separation. In this sense, we recover the behavior of the noninteracting gas. Note however that the magnitude of the interference fluctuations is much enhanced: the $g^{(2)}_I$ drops by a factor of five, which is much larger than the factor of two that we obtain for the Gaussian fluctuations in the noninteracting case of panels (Ic,d). Note that for long separations, the curves in panel IIId) approach each other. This is indeed expected for phase fluctuations that have a linear dispersion.

\begin{figure}[tb]
\includegraphics[width=0.45\textwidth]{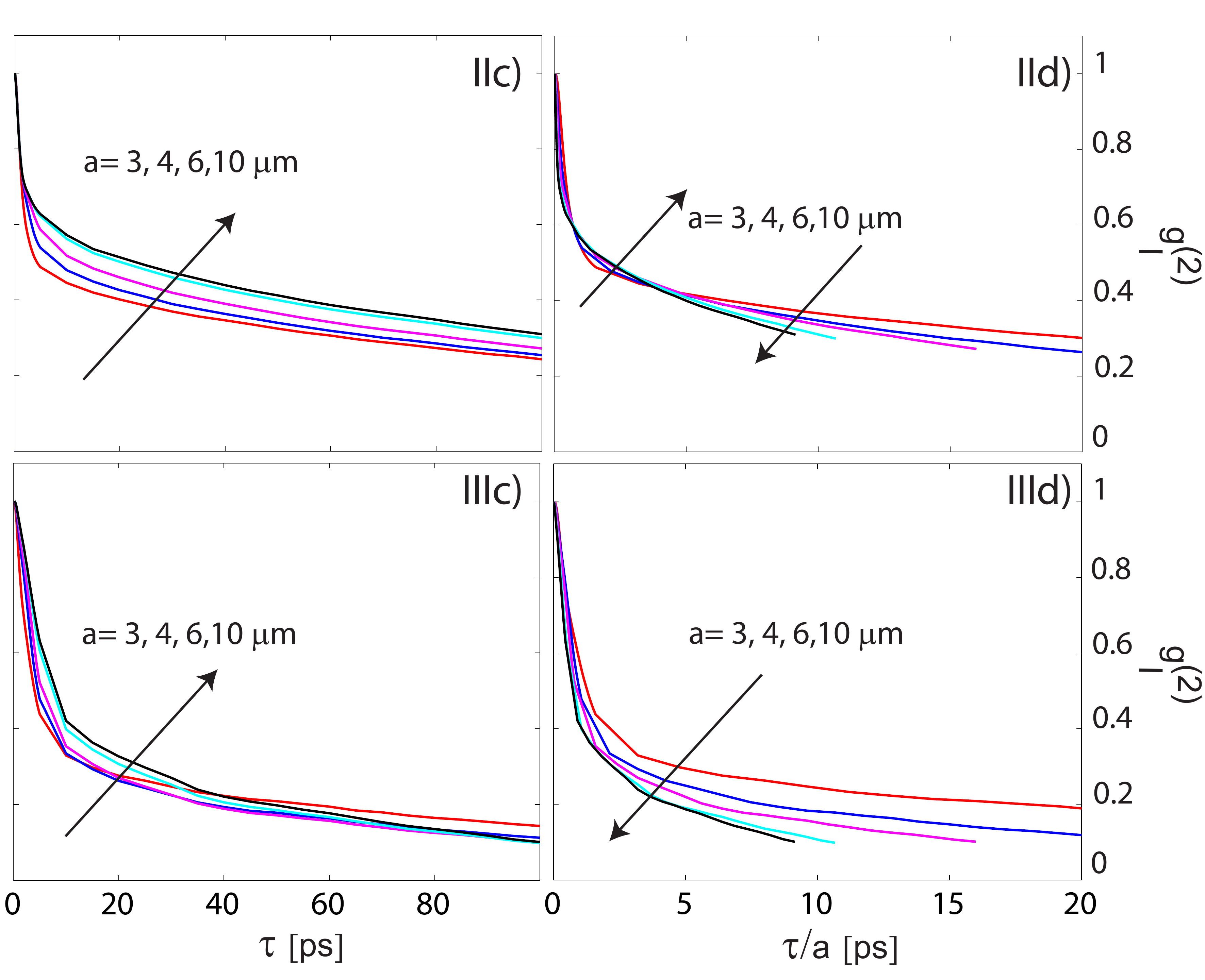}% Here is how to import EPS art
\caption{The same as the corresponding panels in Fig. \ref{fig:results}, but for a finite polariton life time $\gamma/(g n)=0.01$.}
\label{fig:results_finlif}
\end{figure}

%{\em Finite life time --} 
The density fluctuations of the interferrogram can be experimentally measured with standard quantum optical techniques. For this reason, exciton-polariton  quantum fluids in semiconductor microcavities appear as the most natural candidate for studying the phase dynamics in this particular way. A specific aspect of this quantum fluid is the finite life time of the microcavity exciton-polaritons \cite{kasprzak_g2}. Even in the atomic case, a measurement of density fluctuations requires the extraction of atoms from the condensate, limiting the life time of the condensate as well.

Thanks to progress in semiconductor manufacturing technology, the life time of microcavity polaritons has been pushed to the 100 ps regime, equivalent to a line width on the order of 10 $\mu {\rm eV}$, that has to be compared to an interaction energy that can be of the order of 1 eV. We thus expect that in the best microcavities, the effects of a finite life time can be minor. To verify this expectation theoretically, we have repeated our simulations with a finite polariton life time. Numerically, this was implemented by a the inclusion of an imaginary part in the polariton dispersion $\varepsilon(k)$.
Fig. \ref{fig:results_finlif} shows the results of this numerical simulation. The difference between the panels II) and III) is clear, but it is reduced with respect to the case of infinite life time in Fig. \ref{fig:results}. It also appears that that the life time is not long enough in order to observe the diffusive scaling [cf. inset of Fig. \ref{fig:results} IId]. Actually at long times, the quantum fluid enters the noninteracting regime. This is clearly visible in panel IId), where the order of the curves reverses  when $\tau/a \geq  5 {\rm ps/} \mu{\rm m}$.

\section{Conclusions \label{sec:concl}}
In conclusion, we have suggested to study the density fluctuations of the interferrogram to characterize different regimes of a quantum degenerate Bose gas. In particular, a very long correlation time, that shows a ballistic scaling at short distances and a diffusive one at larger distances is observed in a quantum fluid with proliferating vortices. These features mark a distinct contrast with either a noninteracting gas or a Bose-Einstein condensate. 
The numerical evidence for a nontrivial dynamics of these density fluctuations of the interferrogram urges for a deeper theoretical investigations. This could  allow for a more precise characterization of different regimes in quantum turbulence.

\section{Acknowledgements}
This work was performed with financial support from the FWO Odysseus and the UA-LP programs. I acknowledge stimulating discussions with M. Baeten, I. Carusotto, J. Devreese, S. Koghee, F. Manni, and J. Tempere.

\appendix*
\section{Fluctuations in the non-interacting Bose gas}
A Fourier transform of the fields in Eq.\eqref{eq:g2I} yields
\begin{widetext}
\begin{align}
 g^{(2)}_I(\textbf{x},{\bf a},t,&\tau)= 
\int \prod_{j=1}^4  
 \frac{d \textbf{k}_j^D}{(2\pi)^D}
  \exp\{i[(\epsilon_1+\epsilon_2-\epsilon_3-\epsilon_4) t
   +(\epsilon_2-\epsilon_3)\tau-
   (\textbf{k}_1+\textbf{k}_2-\textbf{k}_3-\textbf{k}_4)\cdot\textbf{x}]\} \nonumber \\
  &\times  (1-e^{-i\textbf{k}_1\cdot\textbf{a}})(1-e^{-i\textbf{k}_2\cdot\textbf{a}})(1-e^{i\textbf{k}_3\cdot\textbf{a}})(1-e^{i\textbf{k}_4\cdot\textbf{a}})
\langle \psi^*(\textbf{k}_1)\psi^*(\textbf{k}_2)\psi(\textbf{k}_3)\psi(\textbf{k}_4) \rangle,
\label{eq:g2app}
\end{align}
where $\epsilon_j=\epsilon(k_j)$.
When the phases of the different momentum components are random, the only terms that remain are the usual Wick contractions for which $\textbf{k}_{1,2}=\textbf{k}_{4,3}$ and $\textbf{k}_{1,2}=\textbf{k}_{3,4}$. The first combination gives a contribution that is independent of time, where the second one gives the time dependence. The time dependent part can be written as
\begin{align}
g^{(2)}_I({\bf a} ,t,\tau)-g^{(2)}_I({\bf a} ,t,\tau\rightarrow \infty)= % \nonumber \\
\left|
\int \frac{d^D \textbf{k}}{(2\pi)^D} |\psi(k)|^2 e^{-i\epsilon(k)\tau} 
4 \sin^2\left(\frac{\textbf{k}\cdot\textbf{a}}{2}\right)
\right|^2.
\label{eq:wick}
\end{align}
\end{widetext}
Because for the non-interacting Bose gas the field  $\psi(\xx)-\psi(\xx+\textbf{a})$ has a Gaussian distribution, its density fluctuations show bunching by a factor of two:
$g^{(2)}_I({\bf a} ,t,\tau=0)=2g^{(2)}_I({\bf a} ,t,\tau\rightarrow \infty)$.
With this relation, we obtain from Eq.~\eqref{eq:wick} the expression \eqref{eq:g2I_noint}.

%\bibliography{spontvort}% Produces the bibliography via BibTeX.

\end{document}